\documentclass{article}%
\usepackage{graphicx}
\usepackage{amsmath}
\usepackage{amsfonts}
\usepackage{amssymb}%
\providecommand{\U}[1]{\protect\rule{.1in}{.1in}}

\begin{document}

\author{Antony Valentini\\Augustus College}

\begin{center}
\bigskip

{\LARGE On Galilean and Lorentz invariance in}

\bigskip

{\LARGE pilot-wave dynamics}

\bigskip

\bigskip

Antony Valentini

\bigskip

\bigskip

\textit{Department of Physics, University of Rome `La Sapienza',}

\textit{Piazzale Aldo Moro 2, 00185 Rome, Italy}.\textit{\footnote{Present
address: Theoretical Physics Group, Blackett Laboratory, Imperial College
London, Prince Consort Road, London SW7 2AZ, United Kingdom. email:
a.valentini@imperial.ac.uk}}

\bigskip

\bigskip

\bigskip
\end{center}

It is argued that the natural kinematics of the pilot-wave theory is
Aristotelian. Despite appearances, Galilean invariance is \textit{not} a
fundamental symmetry of the low-energy theory. Instead, it is a fictitious
symmetry that has been artificially imposed. It is concluded that the search
for a Lorentz-invariant extension is physically misguided.

\bigskip

\bigskip

\bigskip

1 Introduction

2 Background

3 The role of kinematics and dynamics

4 Aristotelian kinematics and Aristotelian spacetime

5 Galilean invariance: a fictitious symmetry

6 `Detection' of the natural state

7 Origin of the velocity field

8 Galilean invariance and the quantum potential

9 Remarks

10 Conclusions

\bigskip

\bigskip

\bigskip

Published in: \textit{Physics Letters A} \textbf{228}, 215--222 (1997).

\bigskip

\bigskip\bigskip

\bigskip

\bigskip

\bigskip

\bigskip

\bigskip

\bigskip

\bigskip

\bigskip

\section{Introduction}

The de Broglie--Bohm pilot-wave formulation of quantum theory [1--4] is known
to be mathematically equivalent to standard quantum theory (at least insofar
as the latter is applicable). But with regard to the physical interpretation
of the pilot-wave theory itself, there is as yet no clear consensus.

The dynamics of the theory may be written in a first-order, velocity-based
form [1,5]. For the case of a nonrelativistic system of $N$ particles with
masses $m_{i}$ and three-vector positions $X_{i}(t)$ ($i=1,\ 2,...,\ N$), the
motion of the particles is determined by the de Broglie guidance equation%
\begin{equation}
m_{i}\frac{dX_{i}}{dt}=\nabla_{i}S(X,t)\ .\label{eq1}%
\end{equation}
Here, $S$ is the phase of the wavefunction $\Psi$ (defined on $3N$-dimensional
configuration space), which satisfies the Schr\"{o}dinger equation (units
$\hbar=1$)%
\begin{equation}
i\frac{\partial\Psi}{\partial t}=\sum_{i=1}^{N}-\frac{1}{2m_{i}}\nabla_{i}%
^{2}\Psi+V\Psi\ .\label{eq2}%
\end{equation}
In principle, these equations determine the motion of the particles, given
their initial positions and the initial wavefunction. In practice, the initial
positions are not known. If, for an ensemble of (approximately isolated)
systems with the same wavefunction $\Psi$, the initial distribution of
configurations is given by $P=\left\vert \Psi\right\vert ^{2}$, the
statistical predictions of quantum theory will be accounted for.

It is widely believed that Galilean invariance is a fundamental symmetry of
the above theory. Mathematically, if the coordinates are subjected to a
Galilean transformation%
\[
x_{i}^{\prime}=x_{i}-vt,\ \ \ \ \ t^{\prime}=t
\]
(where, like $v$, each $x_{i}$ is a three-vector), the accompanying
transformation [3]%
\begin{equation}
\Psi^{\prime}=\Psi\exp\left[  i\left(  \frac{1}{2}\sum_{i}m_{i}v^{2}t-\sum
_{i}m_{i}v\cdot x_{i}\right)  \right]  \label{eq3}%
\end{equation}
of the wavefunction leads to the Galilean invariance of (1) and (2), where%

\begin{equation}%
\begin{array}
[c]{l}%
S^{\prime}=S+\frac{1}{2}\sum_{i}m_{i}v^{2}t-\sum_{i}m_{i}v\cdot x_{i}\ ,\\
\ \\
\nabla_{i}^{\prime}S^{\prime}=\nabla_{i}S-m_{i}v\ .
\end{array}
\label{eqn4}%
\end{equation}
It then seems natural to search for a relativistic extension of the theory,
that should be fundamentally Lorentz invariant. This leads to serious
difficulties for the many-body case, as was first noted by de Broglie [1].
Some critics might regard these difficulties as evidence against the
credibility of the whole theory.

In this note we argue that, if the pilot-wave theory is correctly interpreted,
Galilean invariance is \textit{not} a fundamental symmetry of the above
low-energy theory. The search for a Lorentz-invariant extension then seems
misguided. In our view, the difficulties encountered in such a search are no
reflection on the plausibility of the pilot-wave theory. Rather, they show
that the theory is not being interpreted correctly.

\section{Background}

The above dynamics was published by de Broglie in 1928 [1].\footnote{In 1927
de Broglie proposed the theory for a \textit{many}-body system with a
wavefunction in \textit{configuration} space, contrary to many erroneous
claims, made in the literature, that he only considered the one-body theory
with a wavefunction in three-space. (The historical study by Cushing [6] is
also mistaken on this point, as Cushing has recently recognised [7,8].) See
Ref. [1], pp. 118, 119, where de Broglie generalises the one-particle
pilot-wave theory to a many-body system. A full analysis and outline of Ref.
[1] is given elsewhere [9,10].} It was a natural generalisation of his earlier
theory of particle motion, in which velocities are determined by `phase waves'
[11]. For de Broglie, the guidance equation (1) was an expression of the
identity of the principles of Maupertuis and Fermat, and was central to his
view of the dynamics. (From 1925 to 1927 de Broglie tried to derive (1) from
an underlying `double solution' theory. But then he saw that it could simply
be postulated [12], leading to what he called the `pilot-wave theory' based on
(1) and (2).)

Bohm revived the theory in 1952, but in a second-order form based on
acceleration [2]. In Bohm's mechanics, (1) is not regarded as an equation of
motion, but as a restriction imposed on the \textit{initial} momenta [2, p.
170]. This boundary condition happens to be preserved by the time evolution,
which is thought to be generated by a pseudo-Newtonian equation for
acceleration, containing a `quantum potential'. It was suggested by Bohm that
the restriction (1) could be relaxed, leading to corrections to quantum
theory. In contrast, in de Broglie's (mathematically equivalent) dynamics of
1928, the motion is regarded as determined by (1) alone, as de Broglie himself
made very clear [1]. And there is no question of relaxing (1), for it is the
expression of de Broglie's guiding idea -- that the principles of Maupertuis
and Fermat are the same.\footnote{Some authors have erroneously referred to de
Broglie's dynamics, defined by (1) and (2), as `Bohmian mechanics'. This
misnomer not only ignores de Broglie's priority; it also misrepresents the
views of Bohm.}

Bohm showed that the theory -- which de Broglie had applied to interference
and scattering -- could account for the general quantum theory of measurement.
Recently, de Broglie's original first-order dynamics was revived by Bell [5].

The first-order approach, in which (1) is regarded as the basic law of motion
for the system, was adopted by the author [13,14] and by D\"{u}rr et al.
[15,16]. There are some differences, however, which are crucial for the issue
of Galilean and Lorentz invariance.

In the author's view, the pilot wave $\Psi$ should be interpreted as a new
causal agent, more abstract than forces or ordinary fields [14]. This causal
agent is grounded in configuration space -- which is not surprising in a
fundamentally `holistic', nonlocal theory. Heuristically, however, its action
in three-space may be visualised in terms of `Aristotelian forces'. The
`Aristotelian force' $f_{i}=\nabla_{i}S$ on the right-hand-side of (1) is
analogous to the Newtonian force $F_{i}=-\nabla_{i}V$ (where $V$ is the
potential energy) that appears on the right-hand-side of the classical law of
motion%
\begin{equation}
m_{i}\frac{d^{2}X_{i}}{dt^{2}}=-\nabla_{i}V(X,t)\ .\label{eq5}%
\end{equation}
According to (5), the ratio of Newtonian force to mass gives the acceleration.
While according to (1), the ratio of `Aristotelian force' to mass gives the
\textit{velocity}.

Clearly, it is not logically necessary that causal thinking be based on
`Newtonian forces' proportional to the acceleration. Generically, `force'
simply means `agency causing motion (of a specific type)'. And the causal
agency acting on a particle might, in reality, be an `Aristotelian force'
proportional to the velocity (as was in fact widely believed before Galileo).
In our view the physical, causal explanation for why, say, a particle slows
down as it approaches a certain point is simply that the `Aristotelian
potential' $S$ has a turning point there.

In Ref. [14], it was noted that Aristotelian causes are physically
incompatible with the relativity of motion, as remarked by Russell [17]. (This
is partly why they were abandoned in the 17th century.) If one invokes such
causes, there should be a natural state of absolute rest. Thus, in Ref. [14],
only time-independent transformations of three-space coordinates were
considered, with respect to which $\Psi$ is a \textit{scalar}.

In contrast, D\"{u}rr et al. [15,16] do not explicitly invoke such causal
agents, and regard Galilean invariance as a fundamental symmetry of the theory
(so that, as in classical mechanics, uniform motion is relative). Thus $\Psi$
is subject to the transformation law (3). Indeed, D\"{u}rr et al. appeal to
Galilean invariance in order to motivate the choice (1) of velocity field.

This appeal to Galilean invariance has recently been questioned by Brown et
al. [18] who conclude that, indeed, in the context of the pilot-wave theory,
the Aristotelian forces introduced by the author seem more natural than the
usual Newtonian ones. However, the issue of the status of Galilean invariance
was unfortunately not resolved. In particular, it seemed doubtful that a
natural state of rest could be reconciled with the Galilean invariance of (1)
and (2). Brown et al. remark that there appears to be no frame-independent
definition of Aristotelian forces (since these forces will themselves
transform under a Galilean transformation).

We shall now attempt to clarify this potentially very confusing situation.

\section{The role of kinematics and dynamics}

In present physical theories at least, the burden of describing Nature is
shared between kinematics and dynamics. The former defines the structure of
spacetime; the latter accounts for motion within this structure (in so far as
this motion deviates from the `natural', force-free state).

Now the division of labour between kinematics and dynamics cannot be defined
uniquely. For instance, it is usual in classical general relativity to assume
a curved spacetime. But one may equally think in terms of flat spacetime, on
which a `metric field' distorts rods and clocks so as to give the
\textit{appearance} of a curved geometry. In the former approach, a wealth of
phenomena are accounted for by the kinematics of spacetime itself. In the
latter, these same phenomena are accounted for by dynamical influences on flat
spacetime. And it is impossible to say which picture is `true'.

Nevertheless, to say that classical spacetime is flat is usually regarded as a
mistake, for the following reason: The whole purpose of kinematics is to
embrace \textit{universal} features of the dynamics. Any effects that are
found to be independent of the particular material bodies involved are best
assumed to be part of the kinematics.

This point has recently been made particularly clearly by Sonego [19]. The
`zeroth law of mechanics' -- that the behaviour of free bodies is independent
of their mass and composition -- plays a key role in defining the geometry of
spacetime.\footnote{On this basis Sonego argues that, because the spreading of
free wavepackets is mass-dependent, quantum theory alone does not clearly
define a spacetime structure. Some sort of underlying dynamics is necessary.}

Such arguments are, however, usually presented in `operational' terms:
Further, a body is in general assumed to be `free' when it is far away from
other bodies of a specified type (charged bodies in general relativity,
massive bodies in Newtonian theory). Thus it is implicitly assumed that: (i)
Forces have their origin in other bodies; (ii) The effect of these forces
diminishes with distance.

This approach would not be appropriate in the pilot-wave theory. For we are,
at least at present, unable to perform operations with subquantum
trajectories. Further, the origin of forces in the pilot-wave theory is not
other bodies, but the wavefunction itself.\footnote{In (1), the Aristotelian
force acting on a particle depends on the positions of the other particles.
But it does not make sense to regard these other particles as the
\textit{origin} of this force because, unlike in classical mechanics, the
force is not generated by a fixed ($\Psi$-independent) function of relative
particle positions. In this sense the pilot-wave theory is distinctly
non-mechanical, as pointed out (in terms of the quantum potential) by Bohm et
al. [20].} And, of course, these forces do not necessarily diminish with distance.

But it is still possible to define a natural kinematics in the pilot-wave
theory. For while operational arguments sometimes act as guides in the
\textit{construction} of theories, if one is \textit{given} a theory, the
essential point -- that kinematics should embrace universal features of the
dynamics -- may be implemented without recourse to strictly operational
arguments. For the theory itself tells us what `universal' features there are
in Nature (assuming the theory to be true). And once we have identified these,
the most convenient definition of kinematics becomes clear. In other words, if
we are given a theory, a natural definition of kinematics is singled out by
the theory itself.

Note that, as Poincar\'{e} [21] in particular was well aware, it is possible
in principle, in virtually any theory, to adopt virtually any spacetime
structure, provided one adds appropriate compensating dynamical factors. So
the issue is not whether a certain kinematics is \textit{possible}, but
whether it is the most \textit{suitable}.

\section{Aristotelian kinematics and Aristotelian spacetime}

We now show that the Aristotelian dynamics of the pilot-wave theory naturally
selects an Aristotelian kinematics and an Aristotelian spacetime.

In order to define a kinematics, one must first of all find a natural
definition of a `free' system. Otherwise, it will be impossible to find any
`universal' properties of motion at all.

The procedure is a familiar one. In Newtonian mechanics, for instance, forces
are identified as the cause of motion. A system is considered to be `free'
when the right-hand-side of (5) vanishes. One then has a set of `natural
motions'%
\[
X_{i}(t)=v_{i}t+X_{i}(0)\ ,
\]
where $X_{i}(0)$ and $v_{i}$ are arbitrary constants. These trajectories are
completely independent of the masses of the particles. It is therefore
expedient to regard them as features of spacetime itself (that is, as
geodesics associated with an appropriate affine structure). In this way, one
arrives at the kinematics of Galilean spacetime. Uniform motion is the natural
state, and Galilean invariance is a fundamental symmetry.

Similarly, in general relativity one defines a `free' body to be one upon
which no nongravitational forces act. The resulting trajectories are
independent of the body's mass and composition. So again, it makes sense to
regard the trajectories as properties of spacetime itself. Hence the picture
of geodesics in curved spacetime.

What is the natural definition of a `free' system in the pilot-wave theory?
\textit{If} one writes the dynamics in the first-order form given by (1) and
(2), a system can only be regarded as `free' if the `Aristotelian force',
appearing on the right-hand-side of (1), vanishes. One then has a set of
`natural motions'%
\[
X_{i}(t)=X_{i}(0)\ ,
\]
where $X_{i}(0)$ are arbitrary constants. Again, these (rather trivial)
trajectories are completely independent of the masses of the particles. It is
both natural and expedient to regard them as features of spacetime itself.
Thus one arrives at \textit{Aristotelian spacetime}, in which the natural
state of motion is \textit{rest}.

Formally speaking, Aristotelian spacetime may be characterised as a product
$E\times E^{3}$ of a Euclidean time line with Euclidean three-space [22]. The
natural invariance group corresponds to time-independent transformations on
three-space. In particular, the physics is invariant under translations and
rotations in $E^{3}$.

With respect to this natural group, the guiding field $\Psi$ is, as we have
mentioned, a scalar. This ensures that (1) is invariant. (If $V$ depends only
on the distances between particles, (2) will of course also be invariant.)

Note that Aristotelian spacetime $E\times E^{3}$ is a perfectly abstract,
geometrical structure. In principle, no coordinates are required for its description.

\section{Galilean invariance: a fictitious symmetry}

At this point, it might be objected that the equations (1) and (2) are
nevertheless Galilean invariant. But this mathematical symmetry is misleading.

Consider, by way of analogy, the case of classical mechanics. And consider a
transformation of coordinates%
\[
x_{i}^{\prime}=x_{i}-\frac{1}{2}at^{2},\ \ \ \ \ t^{\prime}=t\ ,
\]
to a new frame with an \textit{acceleration} $a$. If one defines a transformation%

\begin{equation}%
\begin{array}
[c]{l}%
V^{\prime}=V-\frac{1}{2}\sum_{i}m_{i}a^{2}t^{2}+\sum_{i}m_{i}a\cdot x_{i}\ ,\\
\ \\
-\nabla_{i}^{\prime}V^{\prime}=-\nabla_{i}V-m_{i}a
\end{array}
\label{eqn6}%
\end{equation}
of the Newtonian potential energy and force, the classical dynamical equations
(5) become invariant. They have the same form in the accelerated frame. The
usual view, of course, is that in (6) one has introduced `fictitious inertial
forces' in the accelerated frame, in order to make Newton's laws
\textit{appear} invariant. By \textit{imposing} an appropriate `transformation
law' on the Newtonian force, one can (artificially) make classical mechanics
invariant under transformations to accelerated frames of reference.

Physically, and mathematically, the transformations (4) and (6) are very
similar. In (6) the Newtonian force $-m_{i}a$ acting on the $i$th particle (in
the accelerating frame) is proportional to the mass; its effect is therefore
independent of the mass. It affects all bodies equally. This universality
betrays it as a fictitious entity that ought to be eliminated by an
appropriate choice of kinematics. Similarly, in (4) the Aristotelian force
$-m_{i}v$ acting on the $i$th particle (in the uniformly moving frame) is
proportional to the mass, has a universal effect, and should be eliminated by
redefining the kinematics.

Thus the supposed `Galilean invariance' of the pilot-wave theory is, in our
view, a first-order analogue of the above \textit{fictitious} invariance of
(second-order) classical mechanics.\footnote{Note that, for a symmetry to be
included in the physical kinematics, it is not enough that it be a mere
mathematical symmetry. For example, it has been known since 1910 that
Maxwell's equations are invariant under a 15-parameter Lie group whose
coordinate transformations include not only Lorentz transformations but also a
scale transformation and a class of nonlinear transformations (where a subset
of the latter corresponds, in the nonrelativistic limit, to a transformation
to an accelerated frame). See, for example, Ref. [23].} Just as the true,
physical invariance group of classical mechanics leaves acceleration and
(Newtonian) force invariant, so the true, physical invariance group of
pilot-wave dynamics leaves velocity and (Aristotelian) force invariant. (This
resolves the difficulty raised by Brown et al.)

D\"{u}rr et al. [15,16] have proposed what is, in effect, a mixture of
first-order (Aristotelian) dynamics with second-order (Galilean) kinematics.
We assert on the basis of the above reasoning that such a mixture is
physically incongruous. An Aristotelian dynamics requires an Aristotelian kinematics.

\section{`Detection' of the natural state}

In the pilot-wave theory of particles, at the fundmental level of an
individual system, the effects of absolute velocity (of the reference frame)
may be compensated for by transforming the wavefunction via (3).

As a result, the natural state of zero velocity cannot be detected. This may
seem peculiar, in view of our assertion that this state exists.

But the situation is similar in classical mechanics. There, strictly speaking,
the natural state of zero acceleration cannot be detected -- without implicit
assumptions about the origin of forces. For by transforming the Newtonian
force according to (6), the effects of absolute acceleration may be cancelled
(that is, the effect of an acceleration of the reference frame on the observed
laws of physics may be compensated for).

Let us be clear about this. Classically, the transformation (6) introduces
what are usually called `fictitious inertial forces'. They are regarded as
fictitious because they appear to have no source. But strictly speaking, the
real existence of such forces cannot be ruled out. They might be generated,
for instance, by acceleration with respect to distant matter. Perhaps they are
real after all. Indeed, even without distant matter, it might simply be that
classical forces really do transform in this way -- just as a magnetic field
may be generated by the Lorentz transformation of an electric field.

Strictly speaking, then, the natural state of unaccelerated motion cannot be
detected in classical mechanics -- unless one \textit{assumes} that real
forces have their origin in nearby bodies. With this assumption (or consensus)
as to what real forces are, a particle far away from other bodies provides a
standard of unaccelerated motion.

There can be no comparable assumption in the pilot-wave theory, where motion
is generated by the wavefunction. One would need a consensus on what the
wavefunction of the universe really is (up to a constant phase), in order to
detect the true frame.

None of this affects our argument, however.

First of all, in both theories, a natural kinematics is in any case singled
out by the dynamics as above, simply by identifying `forces' with the entities
appearing on the right-hand-side of the equations of motion. No further
consensus on the nature of these forces is needed to single out the natural
kinematics, even if practical detection of the natural state should turn out
to be problematic without such a consensus.

Secondly, things are much more clear cut in field theory. There, the theory
not only picks out a natural Aristotelian kinematics (where the natural state
is a static field configuration). The natural state of rest is also singled
out by the detailed behaviour of the field. The (static) vacuum field is not
Lorentz invariant [20]. And if fundamental nonlocality is assumed to define an
absolute simultaneity, the speed of approximately classical electromagnetic
waves (measured by absolutely synchronised clocks) will be isotropic only if
the clocks are at absolute rest [9,14,24]. (If $P\neq\left\vert \Psi
\right\vert ^{2}$, there will be observable instantaneous signals [13]. These
may be used to synchronise distant clocks.)

Even in the case of field theory, no doubt, one could construct other, more
contrived theories with a different (possibly Lorentz-invariant) natural
state. For one may always introduce compensating factors that make unnatural
motions \textit{appear} natural, as seen above in the case of classical
mechanics, where the transformation (6) makes accelerated motion appear
force-free. In practice, however, such compensating factors are easily
recognised to be rather contrived.\footnote{As already noted, the issue is not
whether Lorentz invariance is possible, but whether it is suitable (as is
always the case with kinematics). Despite the difficulties arising from
nonlocality in Minkowski spacetime, one might be able to construct a
high-energy theory with a \textit{fictitious} Lorentz invariance, analogous to
the fictitious Galilean invariance of the low-energy theory. As we have seen,
it would be an error to regard this symmetry as part of the kinematics. (On
the other hand, should attempts to impose Lorentz invariance fail, it would be
a mistake to be alarmed by this. For such attempts have no real motivation,
since even the low-energy theory is not really fundamentally invariant under
boosts.)}

In conclusion, then, so-called `operational detection' of natural motion never
really occurs, in \textit{any} theory, without implicit, simplifying
assumptions.\footnote{The situation is no better in special or general
relativity. The former may be given a Lorentz interpretation, with an absolute
rest frame; while the latter may, as noted, be cast in terms of a flat
spacetime background. Only theoretical expediency singles out the usual
kinematics.} In the end, it is simply the case that the theory itself singles
out a natural, convenient kinematics. And this is as true in the pilot-wave
theory as in any other.

\section{Origin of the velocity field}

It might be thought [15] that Galilean invariance plays an important role,
ensuring that the coefficients $m_{i}$ in (1) are equal to those in (2). But
the structure of the Schr\"{o}dinger equation itself, in \textit{one} frame of
reference, singles out the natural velocity field $\nabla_{i}S/m_{i}$ -- just
as the structure of general relativity singles out a natural set of
trajectories (the geodesics). There is no need to explain why particles follow
precisely the natural trajectories singled out by the theory.\footnote{The
same cannot be said for the equilibrium distribution $P=\left\vert
\Psi\right\vert ^{2}$, even though $\left\vert \Psi\right\vert ^{2}$ is the
natural measure. For (1) and (2) are fundamental laws, while ensemble
distributions are contingent. The distribution $P=\left\vert \Psi\right\vert
^{2}$ should be given a dynamical explanation along the lines of classical
statistical mechanics [19,13,14,25].} (In contrast, in Newtonian physics it is
cogent to ask why inertial and gravitational mass are equal, because there is
nothing in the structure of the theory that prefers this.)

For the Schr\"{o}dinger equation (2) implies the continuity equation%
\[
\frac{\partial\left\vert \Psi\right\vert ^{2}}{\partial t}+\sum_{i=1}%
^{N}\nabla_{i}\cdot\left(  \left\vert \Psi\right\vert ^{2}\frac{\nabla_{i}%
S}{m_{i}}\right)  =0
\]
for the naturally-occurring quantity $\left\vert \Psi\right\vert ^{2}$. Any
function proportional to $\left\vert \Psi\right\vert ^{2}$ is preserved by the
velocity field $\nabla_{i}S/m_{i}$, if and only if the coefficients $m_{i}$
are equal to those in (2). Thus, given that the velocity field is proportional
to $\nabla_{i}S$, the structure of the Schr\"{o}dinger equation singles out
the natural coefficients $1/m_{i}$. There is no need to appeal to Galilean
invariance to fix these coefficients, as done by D\"{u}rr et al. [15].

Note that we are appealing here, not to the empirical conservation of the
equilibrium ensemble distribution $P=\left\vert \Psi\right\vert ^{2}$, but to
mathematical properties of the Schr\"{o}dinger equation for an
\textit{individual} system. For if $\Psi$ is associated with an individual
system, so is the quantity $\left\vert \Psi\right\vert ^{2}$. (Since
probabilities are not fundamental in the pilot-wave theory, the question of
the equality of masses in (1) and (2) must be addressed at the level of
individual dynamics. And at that level, the mathematical structure of the
Schr\"{o}dinger equation does pick out a natural velocity field $\nabla
_{i}S/m_{i}$.)

\section{Galilean invariance and the quantum potential}

To convincingly maintain Galilean invariance as part of the fundamental
symmetry group, one could write the pilot-wave theory in Bohm's second-order,
pseudo-Newtonian form based on the quantum potential. For then, the natural
definition of a `free' system would be analogous to that in classical
mechanics: the total Newtonian force acting on the system (including that
generated by the quantum potential) must vanish. One then obtains the
classical set of natural motions, with arbitrary uniform velocity, leading to
a natural Galilean kinematics. Thus Holland [3] is consistent when he asserts
that Galilean invariance is a fundamental symmetry, for he bases the dynamics
on the quantum potential.

But then things become rather inelegant, and also difficult. The quantum
potential itself is inelegant. The Galilean transformation (3) of the
wavefunction is mathematically peculiar, having no simple geometrical
interpretation. And a Galilean-invariant theory invites attempts at a
Lorentz-invariant extension, leading to enormous complications.

In contrast, the guidance equation (1) is simple and elegant. So is
Aristotelian spacetime $E\times E^{3}$. The wavefunction is simply a scalar.
And there is no need to think of Lorentz invariance as anything other than a
phenomenological feature of the equilibrium distribution $P=\left\vert
\Psi\right\vert ^{2}$ -- as becomes clear in field theory [4,9,14,20,24].

\section{Remarks}

The definition of spacetime structure, and the choice of a true invariance
group, is by no means a merely semantic issue. It has serious ramifications
both for our understanding of the physics and for the subsequent development
of the theory. Field theory may in fact be developed with advantage on
Aristotelian spacetime. Electrodynamics is greatly simplified, since the
time-component of the vector potential (which leads to `ghosts' in standard
QED) makes no appearance at all [14,26]. Similarly, non-Abelian gauge theories
such as QCD have a natural ghost-free formulation, equivalent to standard
field theory written in the temporal gauge [9].

It has been proved [27] that `York time' [28] provides a unique slicing of
curved, classical spacetime, for a closed universe satisfying reasonable
conditions. York time may then be regarded as an absolute cosmic time [29,30].
It is natural to identify York time with our Aristotelian time, and the
associated curved spacelike slices with \textit{curved} Aristotelian
three-space [9,26]. One may then identify gravitation as a curvature of
Aristotelian spacetime $E\times E^{3}$, rather than as a curvature of
Minkowski spacetime $M_{4}$. Since the former has a definite foliation built
into it, the notorious `problem of time' in quantum gravity is eliminated from
the outset. (Technical problems remain, however.)

An Aristotelian kinematics, with an absolute slicing of spacetime, is in fact
naturally singled out by the pilot-wave dynamics of gravity. For the momentum
canonically conjugate to the three-metric is given essentially by the
extrinsic curvature tensor -- which tells how the three-geometry is embedded
in spacetime. Just as the momentum of a particle is determined by its position
and wavefunction in the low-energy theory, so the embedding of a spacelike
slice in spacetime should be determined by its three-geometry and
wavefunction.\footnote{It is therefore wrong to assume, as many authors do, a
guidance equation for the three-metric with arbitrary lapse and shift
functions (that is, with arbitrary spacetime slicing). For a full discussion
see Ref. [9].}

\section{Conclusions}

The second-order structure of classical mechanics defines a physical, Galilean
spacetime with a natural state of zero acceleration. Similarly, the
first-order structure of pilot-wave dynamics defines an Aristotelian
spacetime, with a natural state of zero velocity.

To impose Galilean invariance on the pilot-wave theory is like imposing, on
Newtonian mechanics, an invariance under transformations to uniformly
accelerated frames. The Galilean transformation (3) of $\Psi$ amounts to the
introduction of fictitious inertial (Aristotelian) forces.

Despite appearances, then, Galilean invariance is \textit{not} a fundamental
symmetry of the low-energy pilot-wave theory. There is then no reason to
impose Lorentz invariance in the high-energy domain.

Some authors, following Bell, have portrayed the current situation as a sort
of two-horse race for fundamental Lorentz invariance, the contestants being
the pilot-wave and dynamical-reduction theories. From the above perspective,
this is quite misguided. For in the pilot-wave theory, uniform motion is not
relative -- so the `problem' of finding a Lorentz-invariant extension simply
does not arise. Whether the theory of dynamical reduction is also able to
circumvent this problem remains to be seen. (Perhaps the reduction mechanism
could be shown to single out a natural state of rest.)

\bigskip

\bigskip

\bigskip

\begin{center}
\textbf{Acknowledgement}
\end{center}

It is a pleasure to thank Professor M. Cini for his kind support, and Dr. S.
Sonego for valuable discussions, suggestions and correspondence. I am also
grateful to Dr. H.R. Brown and Dr. J. Butterfield for comments on the manuscript.

\begin{center}
\textbf{References}
\end{center}

[1] L. de Broglie, in: Electrons et photons: rapports et discussions du
cinqui\`{e}me conseil de physique, eds. J. Bordet et al. (Gauthier-Villars,
Paris, 1928).

[2] D. Bohm, Phys. Rev. 85 (1952) 166, 180.

[3] P.R. Holland, The quantum theory of motion: an account of the de
Broglie--Bohm causal interpretation of quantum mechanics (Cambridge Univ.
Press, Cambridge, 1993).

[4] D. Bohm and B.J. Hiley, The undivided universe: an ontological
interpretation of quantum theory (Routledge, London, 1993).

[5] J.S. Bell, Speakable and unspeakable in quantum mechanics (Cambridge Univ.
Press, Cambridge, 1987).

[6] J.T. Cushing, Quantum mechanics: historical contingency and the Copenhagen
hegemony (Univ. of Chicago Press, Chicago, 1994).

[7] J.T. Cushing, in: Experimental metaphysics: quantum mechanical studies in
honor of Abner Shimony, eds. R.S. Cohen and J. Stachel (Kluwer, Dordrecht, 1996).

[8] J.T. Cushing, in: Bohmian mechanics and quantum theory: an appraisal, eds.
J.T. Cushing, A. Fine and S. Goldstein (Kluwer, Dordrecht, 1996).

[9] A. Valentini, On the pilot-wave theory of classical, quantum and
subquantum physics (Springer, Berlin), to be published.

[10] A. Valentini, The early history of Louis de Broglie's pilot-wave
dynamics, in preparation.

[11] L. de Broglie, Recherches sur la th\'{e}orie des quanta, doctoral thesis
(1924), Facult\'{e} des Sciences, Paris. [Ann. de Phys. 3 (1925) 22; English
translation (extracts): G. Ludwig, Wave mechanics (Pergamon, Oxford, 1968).]

[12] L. de Broglie, J. Phys. (Paris) 8 (1927) 225.

[13] A. Valentini, Phys. Lett. A 156 (1991) 5; 158 (1991) 1.

[14] A. Valentini, On the pilot-wave theory of classical, quantum and
subquantum physics, doctoral thesis (1992), International School for Advanced
Studies, Trieste [http://www.sissa.it/ap/PhD/Theses/valentini.pdf].

[15] D. D\"{u}rr, S. Goldstein and N. Zanghi, J. Stat. Phys. 67 (1992) 843.

[16] D. D\"{u}rr, S. Goldstein and N. Zanghi, Phys. Lett. A 172 (1992) 6.

[17] B. Russell, History of western philosophy (Allen \& Unwin, London, 1946).

[18] H.R. Brown, A. Elby and R. Weingard, in: Bohmian mechanics and quantum
theory: an appraisal, eds. J.T. Cushing, A. Fine and S. Goldstein (Kluwer,
Dordrecht, 1996).

[19] S. Sonego, Phys. Lett. A 208 (1995) 1.

[20] D. Bohm, B.J. Hiley and P.N. Kaloyerou, Phys. Rep. 144 (1987) 321.

[21] H. Poincar\'{e}, La science et l'hypoth\`{e}se (Paris, 1902) [English
translation: Science and hypothesis (Dover, New York, 1952)].

[22] R. Penrose, in: Battelle rencontres, eds. C.M. De Witt and J.A. Wheeler
(Benjamin, New York, 1968).

[23] A.O. Barut, Electrodynamics and classical theory of fields and particles
(Dover, New York, 1980).

[24] D. Bohm and B.J. Hiley, Found. Phys. 14 (1984) 255.

[25] D. Bohm, Phys. Rev. 89 (1953) 458.

[26] A. Valentini, in: Bohmian mechanics and quantum theory; an appraisal,
eds. J.T. Cushing, A. Fine and S. Goldstein (Kluwer, Dordrecht, 1996).

[27] J.E. Marsden and F.J. Tipler, Phys. Rep. 66 (1980) 109.

[28] J.W. York, Phys. Rev. Lett. 28 (1972) 1082.

[29] J. Isenberg and J.A. Wheeler, in: Relativity, quanta, and cosmology, Vol.
I, eds. M. Pantaleo and F. de Finis (Johnson Reprint Corporation, New York, 1979).

[30] A. Qadir and J.A. Wheeler, in: From SU(3) to gravity, eds. E. Gotsman and
G. Tauber (Cambridge Univ. Press, Cambridge, 1985).

\end{document}